\documentclass[aps,pre,superscriptaddress,showpacs,notitlepage,a4,twocolumn]{revtex4-2}
\usepackage{amssymb,amsmath}
\usepackage{graphicx}
\usepackage{mathtools}
\usepackage[export]{adjustbox}
\usepackage{overpic}
\usepackage[colorlinks=true,
 linkcolor=black,
 urlcolor=blue,
 citecolor=blue]{hyperref}

\usepackage{mathscinet}
\usepackage[normalem]{ulem}

\usepackage{float}
\usepackage{dcolumn}
\usepackage{bm}
\usepackage{bigints}
\usepackage{xcolor}
\usepackage{appendix}
\usepackage{graphicx}
\usepackage{dcolumn}
\usepackage{bm}
\usepackage{hyperref}
\usepackage{xcolor}
\usepackage{comment}
\usepackage{geometry}
\usepackage{multirow}
\usepackage{graphicx} 
\usepackage{amsmath}
\usepackage{enumitem}

\newcommand{\Tc}{T_\text{c}}

\begin{document}

\title{Barrier crossing in a two-state system: Effect of bias and stochastic fields}
\author{Sara Oliver-Bonafoux}
\affiliation{Instituto de Fisica Interdisciplinar y Sistemas Complejos IFISC (CSIC-UIB), Campus UIB, 07122 Palma de Mallorca, Spain}
\author{Ra\'ul Toral}
\affiliation{Instituto de Fisica Interdisciplinar y Sistemas Complejos IFISC (CSIC-UIB), Campus UIB, 07122 Palma de Mallorca, Spain}
\author{Amitabha Chakrabarti}
\affiliation{Department of Physics, Kansas State University, Manhattan, KS 66506, USA}

\date{\today}

\begin{abstract}
We study barrier crossing in a two-state system, namely the kinetic Ising model, in the presence of a weak bias field and spatially homogeneous, but time-dependent, Gaussian random fields. We find that the bias field determines the location of the dominant maxima of the probability distribution function of the magnetization, whereas the noise intensity controls their sharpness and stability of the distribution. A moderate stochastic field lowers the effective energy barrier and facilitates transitions between ordered states, while strong noise induces broad distributions and significant backflow, which reduces directional selectivity. Our results suggest that efficient barrier crossing requires a balanced combination of moderate stochastic driving and controlled bias.
\end{abstract}

\maketitle

\section{Introduction}

Control of chemical reaction rates has always been an important goal in chemistry and biology~\cite{hromadova2014stochastic}. The ability to accelerate reactions without inducing adverse side effects~\cite{wang2020boosted} would represent a major advance in areas such as interfacial science, biochemistry, synthetic 
chemistry, and catalysis. In general, reaction rates depend exponentially on the height of the free-energy barrier separating the initial and final states~\cite{panta2023quantitatively}. Traditional approaches to enhance reaction rates therefore rely either on increasing the temperature, thereby promoting thermal activation over the barrier, or on employing catalysts that modify the reaction pathway and reduce the activation energy.

More recently, an alternative strategy has emerged in which externally applied stochastic electric fields are used to accelerate chemical and biochemical processes~\cite{martinez2013effective, paneru2023bona}. In such schemes, noise-driven electric fields effectively increase the kinetic energy of charged species, facilitating barrier crossing without raising the thermodynamic temperature of the surrounding medium. Because the bulk temperature remains unchanged, this approach avoids thermal damage and is therefore particularly attractive for biological systems. Some experimental approaches apply a weakly fluctuating field (a bias field)~\cite{Flanders} in addition to a rapidly fluctuating field, and analyze the amplification of the effect of the weak bias field. This amplification is closely related to stochastic resonance phenomena~\cite{gammaitoni1998stochastic}.

Motivated by these developments, we previously investigated a minimal theoretical framework for studying barrier crossing in the presence of driving by stochastic fields~\cite{paper1}. We considered a two-state system represented by a kinetic Ising model coupled to a heat bath, where the two magnetized states correspond to the minima of an effective double-well potential separated by a barrier. In this description, thermal fluctuations originate from the heat bath, while additional nonequilibrium fluctuations are introduced through a spatially uniform, but time-dependent, zero-mean random magnetic field acting on all spins. The stochastic magnetic field thus plays the role of an external noise source that modulates the effective barrier between the two magnetized states.

The kinetic Ising model under time-dependent magnetic fields has been extensively studied in the context of dynamic phase transitions~\cite{Yuksel:2022}. Periodically driven systems were shown to exhibit symmetry-breaking transitions of the time-averaged magnetization, including tricritical behavior at sufficiently large driving amplitudes~\makebox{\cite{tome1990dynamic,lo1990ising,acharyya1995response,acharyya1999nonequilibrium}}. When the magnetic field varies randomly in time, continuous transitions between ordered and disordered phases were reported, based primarily on analyses of the dynamic order parameter~\cite{Acharyya1998}. Additional studies have examined scaling behavior and critical exponents in related setups. For example, Ghosh and Chakrabarti analyzed fluctuations of the time-averaged magnetization under binary stochastic driving and performed finite-size scaling analyses~\cite{Ghosh_2013}, while Li and Wang investigated critical temperatures, hysteresis properties, and finite-size scaling in a kinetic Ising model with randomness~\cite{Li_2024}.

In most of these investigations, the central objective was to characterize phase transitions and determine universality classes through the behavior of the time-averaged magnetization. Our previous work, in contrast, shifted the emphasis from the analysis of the order parameter and its behavior at the transition points to the full probability distribution of the magnetization. This approach revealed that the presence of a stochastic magnetic field fundamentally alters the dynamical structure of the system, leading to regimes in which barrier crossing between preferred states is either enhanced or suppressed depending on temperature and noise strength.

In the present work we further develop this kinetic perspective. We introduce a weak steady bias field that makes one of the two magnetized states energetically favorable, thereby simulating a preferred reaction product in a two-state system. We then investigate how the interplay between thermal fluctuations, stochastic magnetic fields, and the applied bias controls the barrier-crossing time between states. Our primary goal is to determine under what conditions stochastic driving accelerates transitions and how this acceleration depends on temperature and noise intensity.

Although most studies of kinetic Ising systems under time-dependent magnetic fields have been theoretical or computational, recent experiments have demonstrated that nonequilibrium dynamic phase transitions and associated fluctuation phenomena can be observed and characterized in driven magnetic systems. In particular, experiments on ferromagnetic thin films under oscillatory fields have revealed dynamic ordering, critical behavior, and fluctuation regimes without direct equilibrium analogues~\cite{MarinRamirez2020,Quintana2023}. Although the stochastic driving considered here differs from the periodic forcing employed in those studies, these experiments illustrate that externally driven magnetic systems can now be explored with sufficient control to access nonequilibrium fluctuation phenomena. By framing the kinetic Ising model as a minimal representation of a bistable reaction coordinate, our aim is to provide theoretical insight into stochastic-field-assisted barrier crossing and to suggest possible routes for experimentally probing related acceleration phenomena in externally driven systems.

The rest of the paper is organized as follows: In Sec.~\ref{sec:model}, we introduce the model and describe the numerical simulation scheme. In Sec.~\ref{sec:results} we present the results. In Sec.~\ref{sec:results_nobias} we summarize the main findings of our previous work~\cite{paper1} in the absence of a bias field. In Sec.~\ref{sec:results_bias} we discuss the effects of introducing a bias field, and in Sec.~\ref{sec:results_jumptime} we analyze barrier escape times. Finally, conclusions and a discussion of the results are given in Sec.~\ref{sec:conclusions}.

\section{Model and simulation scheme}\label{sec:model}

We consider a kinetic Ising model defined on a two-dimensional square lattice of linear size $L$ with periodic boundary conditions. Each lattice site $i$ carries a spin variable $s_i=\pm1$, and the total number of spins is $N=L^2$. The Hamiltonian reads
\begin{equation}
\mathcal{H}(t) = - J \sum_{\langle i,j\rangle} s_i s_j-h(t) \sum_{i=1}^N s_i ,
\label{eq:Hamiltonian}
\end{equation}
where $\langle i,j\rangle$ indicates the set of nearest neighbors in the square lattice; $J>0$ is the ferromagnetic coupling, and $h(t)$ is a spatially uniform, time-dependent magnetic field.

The system is coupled to a thermal bath at temperature $T$ and evolves according to single-spin heat-bath dynamics~\cite{Newman2023}. At each elementary update, a randomly selected spin is reassigned according to
\begin{eqnarray}
P_t(s_i = +1) 
&=& \frac{1}{1 + \exp\left[-2\beta b_i(t)\right]}, \\
P_t(s_i = -1) &=& 1 - P_t(s_i = +1),
\end{eqnarray}
where $\beta = 1/(kT)$ and the local field acting on spin $i$ is
\begin{equation}
b_i(t) = J \sum_{\mu} s_{i_\mu} + h(t),
\end{equation}
with the sum extending over the four nearest neighbors of site $i$. Throughout this work we set $J=1$ and $k=1$, thereby fixing the temperature scale. Time is measured in Monte Carlo steps (MCS), with one MCS corresponding to $N$ single-spin update attempts.

The time-dependent magnetic field is split as
\begin{equation}\label{eq:magneticfield_contributions}
 h(t)=h_r(t)+H,
\end{equation}
where $h_r(t)$ denotes the stochastic contribution and $H$ represents the steady bias field. The random field $h_r(t)$ follows a Gaussian distribution with zero mean and variance $2D$, where the field (or noise) intensity $D$ controls the strength of the fluctuations. The random field remains constant during each MCS, and then is redrawn independently. Thus, in addition to thermal noise from the heat bath, the system experiences an external stochastic driving that modulates the effective energy landscape in time.

To characterize the macroscopic behavior of the system, we monitor the instantaneous magnetization per spin,
\begin{equation}
m(t) = \frac{1}{N} \sum_{i=1}^{N} s_i(t).
\end{equation}
After discarding an initial transient of duration $t_0$~MCS, measurements are taken at every MCS over a period of $t_{\mathrm{obs}}$~MCS. These data are used to construct the stationary probability distribution of magnetization $P(m)$, from which we obtain the time-averaged magnetization as
\begin{equation}
Q = \frac{1}{t_{\mathrm{obs}}}
\sum_{t=t_0}^{t_0+t_{\mathrm{obs}}} m(t)
 = \sum_m m P(m),
\end{equation}
which serves as a dynamic order parameter. While~$Q$ captures symmetry breaking, our analysis relies primarily on the full distribution $P(m)$, which provides direct insight into metastability and barrier-crossing processes. 

All results presented in this work are obtained from numerical simulations of the heat-bath dynamics described above, and the numerical details specific to each figure are given in the corresponding captions.

\section{Results}\label{sec:results}

\subsection{Results in the absence of a bias field}\label{sec:results_nobias}

In our previous work~\cite{paper1}, we analyzed the model in the absence of a bias field, which corresponds to setting $H = 0$ in Eq.~\eqref{eq:magneticfield_contributions}. For completeness, we briefly summarize the main results here. 

\begin{figure}[h!]
 \centering
 \includegraphics[width=1.0\linewidth]{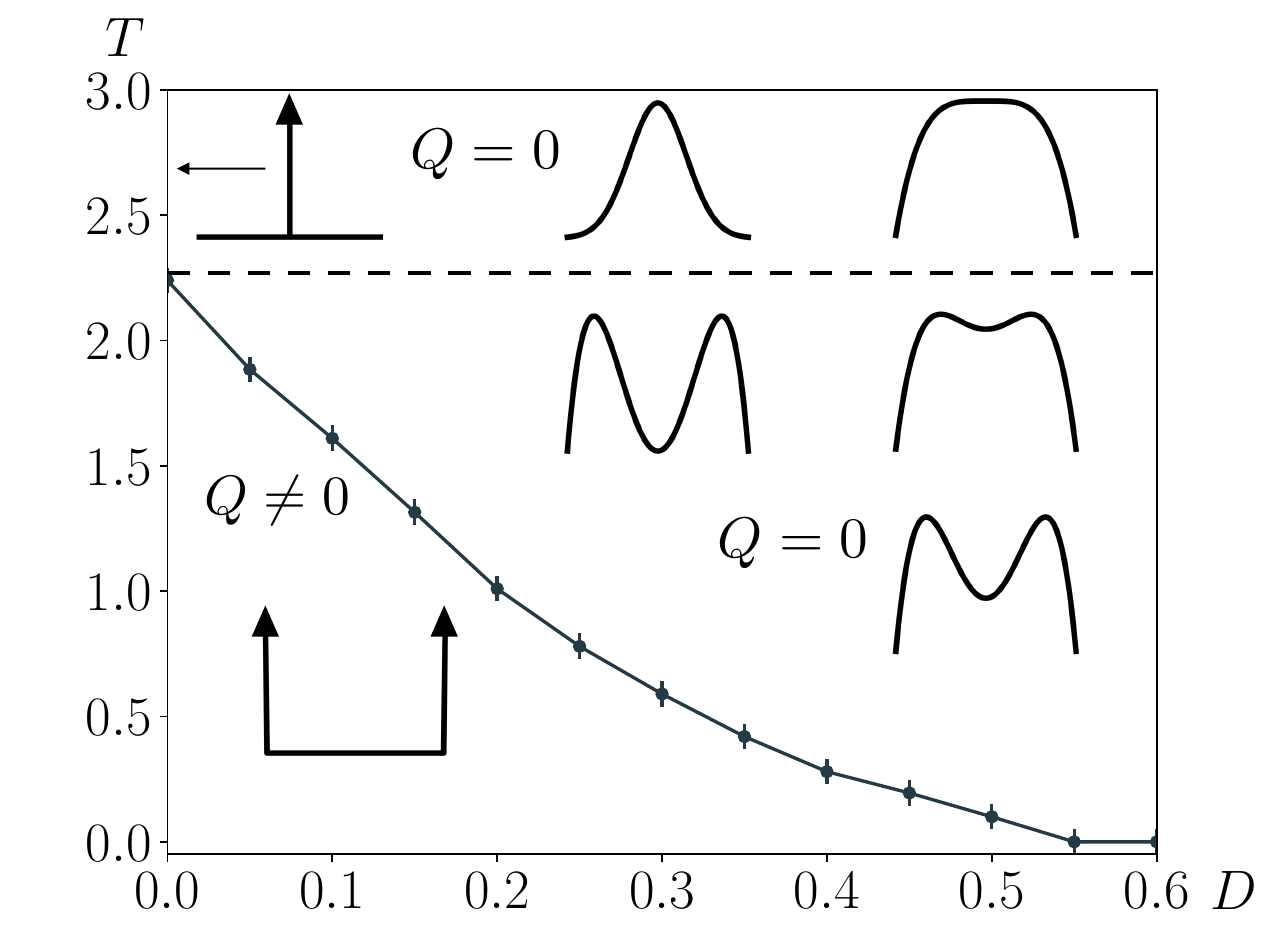}
 \caption{Phase diagram of the model defined by Eq.~\eqref{eq:Hamiltonian} in the absence of a bias field ($H=0$) in the $(D,T)$ plane, where $T$ is the temperature and $2D$ the variance of the Gaussian random field. The sketches illustrate the qualitative shape of the stationary magnetization distribution $P(m)$ in the thermodynamic limit and the corresponding value of the time-averaged magnetization $Q$ in each region. The horizontal dashed line marks the critical temperature of the field-free Ising model, $\Tc\approx2.269$. The solid line separates the broad-ferromagnetic and ferromagnetic phases. It is obtained numerically by identifying, for several values of $D$, the temperature below which no switching events are observed during an observation time $t_{\mathrm{obs}}=10^8$ MCS. The figure has been adapted from~\cite{paper1}.}
 \label{fig:PhaseDiagramH0}
\end{figure}

The phase diagram of the model as a function of stochastic intensity $D$ and temperature $T$ is shown in Fig.~\ref{fig:PhaseDiagramH0} for $H = 0$. For $D=0$, the system displays the well-known second-order phase transition between a paramagnetic phase (characterized by a zero time-averaged magnetization, $Q = 0$) and a ferromagnetic phase ($Q \ne 0$), at the critical temperature $\Tc=2.269\dots$~\cite{Onsager1944crystal,Baxter2011}. The transition is symmetry-breaking: one of the two possible ferromagnetic states, $Q>0$ or $Q<0$, is dynamically selected as $T$ decreases below $\Tc$. In the thermodynamic limit, the magnetization distribution becomes sharply concentrated around the corresponding macroscopic equilibrium values in each phase.

For $D>0$, the stochastic field qualitatively modifies the equilibrium behavior, giving rise to three dynamical regimes in the phase diagram in Fig.~\ref{fig:PhaseDiagramH0}.

At high temperatures, $T>\Tc$, the system exhibits a \emph{broad-paramagnetic} phase. In this regime, the probability distribution of the magnetization, $P(m)$, has a single maximum around $m=0$ (and, therefore, $Q = 0$) but retains a finite width even in the thermodynamic limit. 

At intermediate temperatures, \makebox{$T_0(D)<T<\Tc$}, a \emph{broad-ferromagnetic} phase emerges. In this regime, $P(m)$ develops two symmetric maxima at nonzero magnetization values. However, the system continues to switch dynamically between the two ordered states on accessible time scales, so that the global $\mathbb{Z}_2$ symmetry is restored over long times and the time-averaged magnetization vanishes ($Q = 0$). In terms of an effective free-energy landscape, the stochastic field continuously modulates the barrier separating the two minima, allowing repeated barrier-crossing events even in large systems.

The transition between the broad-paramagnetic and broad-ferromagnetic phases is a noise-induced transition in the sense of Horsthemke and Lefever~\cite{1984Horsthemke} and of related works on stochastic symmetry breaking~\cite{BPT:1994,BPTK:1997,Toral:2011}. For sufficiently weak field intensity $D$, this transition occurs at the equilibrium critical temperature of the two-dimensional Ising model.

At lower temperatures, $T<T_0(D)$, a second transition separates the broad-ferromagnetic regime from a low-temperature regime that behaves, for all practical purposes, as a ferromagnetic phase with $Q \ne 0$. In this regime, one of the two symmetry-related ordered states is dynamically selected and switching between them becomes exceedingly rare. The defining signature of this transition is the behavior of the escape time $\tau$ required for the system to leave a fully ordered configuration. As the temperature decreases (or the field intensity is reduced), $\tau$ grows rapidly and eventually exceeds the observation time $t_{\mathrm{obs}}$, so that no switching events are detected within accessible simulation times and barrier crossing becomes effectively frozen. Whether $\tau$ truly diverges in the thermodynamic limit or instead remains finite but extremely large cannot be determined from our simulations. Although the magnetization changes discontinuously across this boundary, the observed behavior does not display the standard characteristics of an equilibrium first-order transition~\cite{Vollmayr1993finite}. Rather, it is controlled by the rapid growth of a characteristic dynamical time scale.

As discussed in~\cite{paper1}, the zero-mean stochastic magnetic field therefore converts the conventional symmetry-breaking transition of the Ising model into a noise-controlled scenario in which barrier crossing between metastable states can be enhanced without necessarily shifting the equilibrium critical temperature.

\subsection{Effect of a steady bias field}\label{sec:results_bias}

In this section, we extend our previous analysis in~\cite{paper1} by introducing a constant bias field \(H \neq 0\), allowing us to study the switching dynamics between the two magnetized states under a constant external driving. Specifically, our primary goal is to determine the conditions under which the stochastic field accelerates these transitions, and how this noise-induced acceleration depends on temperature, noise intensity, and system size in the presence of the bias. From the viewpoint of reaction kinetics, the magnetization acts as a collective reaction coordinate in a bistable landscape. Barrier crossing between the two magnetized states is analogous to a chemical reaction event. In this sense, the model provides a minimal framework to study noise-induced enhancement of barrier crossing in two-state systems, with potential relevance to externally driven chemical and biochemical reactions.

\begin{figure}[h!]
 \centering \includegraphics[width=1.0\linewidth]{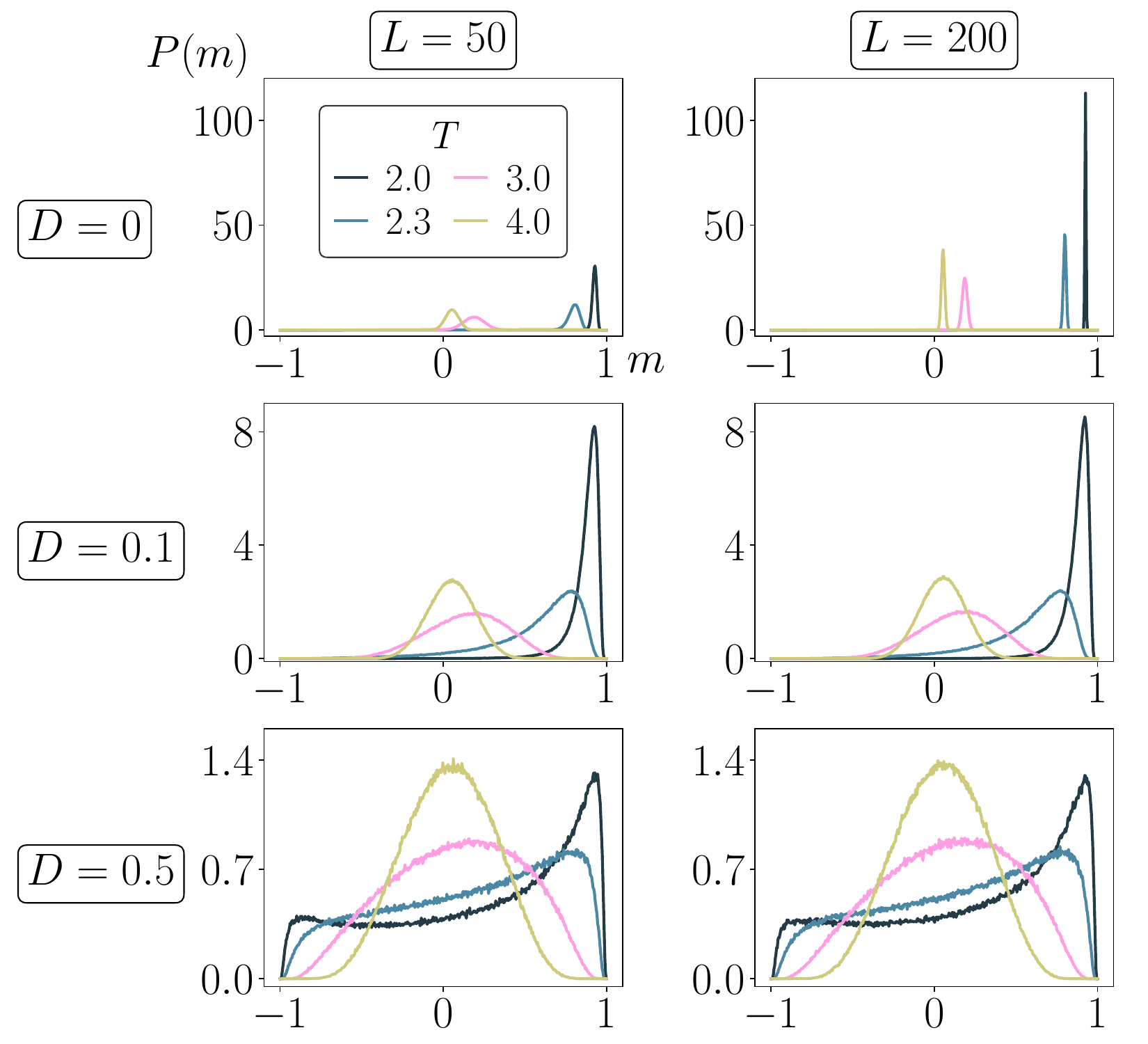}
 \caption{Probability distribution of the magnetization, $P(m)$, for a fixed bias field $H=0.05$ and varying random field intensities ($D = 0, 0.1, 0.5$), and for two system sizes ($L=50,200$). Note that graphs in the same row share the same vertical scale, indicating that, except for the zero-variance case $D=0$, the distributions exhibit no  significant dependence on the system size $L$. This behavior is further analyzed in Fig.~\ref{fig:HistogramsH005T20}.  Results correspond to $10^6$ measurements after a thermalization period of $10^4$ MCS. In the simulations, we take measurements of the magnetization over an observation period of $t_\text{obs}=10^6$ MCS after a thermalization time $t_0=10^4$ MCS. Histograms are constructed using 500 bins in the interval $m\in[-1,1]$.}
 \label{fig:HistogramsH005}
\end{figure}

We first analyze how the phase diagram behavior is modified by the addition of a bias field~$H$. In Fig.~\ref{fig:HistogramsH005}, we show the magnetization distribution $P(m)$ for a fixed bias field $H=0.05$, two system sizes ($L=50$ and 200), and four representative temperatures ($T=2.0$, $2.3$, $3.0$, and $4.0$). For each temperature, several noise intensities ($D=0,0.1,0.5$) are considered, allowing us to explore the combined effects of thermal fluctuations, stochastic driving, and bias. The chosen temperatures probe distinct regimes of the equilibrium Ising model: $T=2.0$ lies below the two-dimensional critical temperature \makebox{$\Tc \approx 2.269$}, $T=2.3$ is slightly above $\Tc$, and $T=3.0$ and $T=4.0$ lie well within the paramagnetic phase. This range allows us to assess how a weak bias modifies the magnetization distribution both near and far from criticality, as well as how the stochastic noise competes with thermal fluctuations in shaping the effective barrier.

For the equilibrium case $D=0$ (see the upper panels in Fig.~\ref{fig:HistogramsH005}), the effect of the bias is straightforward. At $T=2.0$, $P(m)$ displays a single sharp maximum at positive magnetization, corresponding to the ordered state aligned with the field. This peak becomes increasingly sharp with system size, consistent with stabilization of the ordered phase in the thermodynamic limit. At $T=2.3$, the distribution remains biased toward positive $m$, while finite-size effects become more apparent, leading to broader distributions for smaller $L$ due to enhanced critical fluctuations. For $T=3.0$ and $T=4.0$, the maximum shifts to small positive magnetization values, consistent with linear response in the paramagnetic regime, and narrows with increasing system size.

For moderate noise intensity $D=0.1$ (see the middle panels in Fig.~\ref{fig:HistogramsH005}), the distributions remain biased toward positive $m$ due to the bias field, but become noticeably broader at all temperatures. Even at $T=2.0$, the peak does not sharpen significantly with increasing $L$, indicating that stochastic driving weakens the effective barrier between the ordered states. Near and above~$\Tc$, the distributions are substantially broader than in the equilibrium case and exhibit reduced finite-size effects. Thus, moderate stochastic driving enhances fluctuations and promotes barrier crossing while preserving the bias-induced symmetry breaking.

For strong noise intensity $D=0.5$ (see the lower panels in Fig.~\ref{fig:HistogramsH005}), the dynamics is dominated by stochastic driving. In this regime, the critical temperature separating the broad-ferromagnetic and ferromagnetic phases is close to zero, as indicated by the phase diagram in Fig.~\ref{fig:PhaseDiagramH0}. At high temperatures ($T=3.0,4.0$), the distributions are broad and centered around $m\simeq 0$, reflecting fluctuations that overwhelm the bias field. At lower temperatures, such as $T=2.0$, a maximum near $m=1$ develops, but it remains broad and coexists with a weak secondary peak near $m=-1$. This persistent \emph{backflow} into the less populated basin shows that strong stochastic driving does not simply produce one-way transitions toward the preferred state selected by the bias; instead, it continually redistributes probability between the two minima.
 
\begin{figure}[h!]
 \centering
 \includegraphics[width=1.0\linewidth]{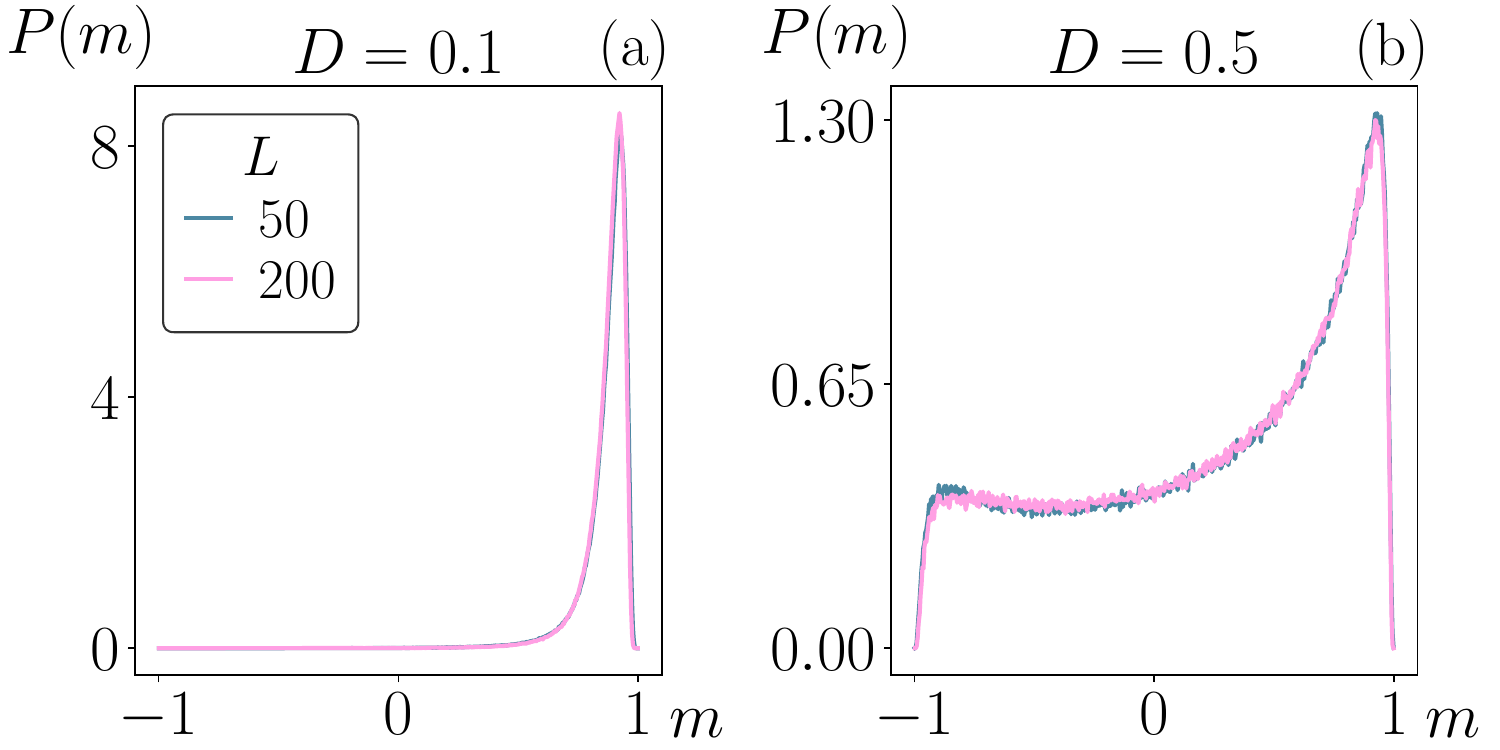}
 \caption{Probability distribution of the magnetization, $P(m)$, for a bias field $H = 0.05$ and temperature \makebox{$T = 2.0$} for two system sizes ($L=50,200$) and two random field intensities: (a) $D = 0.1$ and (b) $D = 0.5$. These results confirm the lack of dependence on the system size observed in Fig.~\ref{fig:HistogramsH005}. In the numerical simulations, the thermalization and measurement times have been set as in Fig.~\ref{fig:HistogramsH005}.}
 \label{fig:HistogramsH005T20}
\end{figure}

The weak finite-size dependence observed for $D>0$ is more clearly illustrated in Fig.~\ref{fig:HistogramsH005T20} for $T=2.0$. For both $D=0.1$ and $D=0.5$, the distributions for $L=50$ and $L=200$ nearly overlap, in contrast with the equilibrium case (see the upper panels in Fig.~\ref{fig:HistogramsH005}). Stochastic driving therefore prevents the progressive sharpening of $P(m)$ with system size and maintains macroscopic fluctuations even in relatively large systems.

\subsection{Barrier escape times}\label{sec:results_jumptime}

To quantify barrier crossing more directly, we analyze the first-passage (or escape) time $\tau$. We initialize the system in the fully ordered configuration with magnetization $m(0)= -1$ (for $H\ge 0$), and define $\tau$ as the first time at which $m(t)$ reaches zero,
\begin{equation}
\tau = \inf\{t>0:\; m(t)=0\}.
\end{equation}
This quantity provides an operational measure of the time required for the system to escape from the unfavorable basin and reach the vicinity of the barrier separating the two magnetization states. By analyzing $\tau$ as a function of temperature, noise intensity, bias strength and system size, we can directly quantify under which conditions stochastic driving facilitates barrier crossing and how this process is influenced by the competing sources of noise. 

\begin{figure}[h!]
 \centering \includegraphics[width=1.0\linewidth]{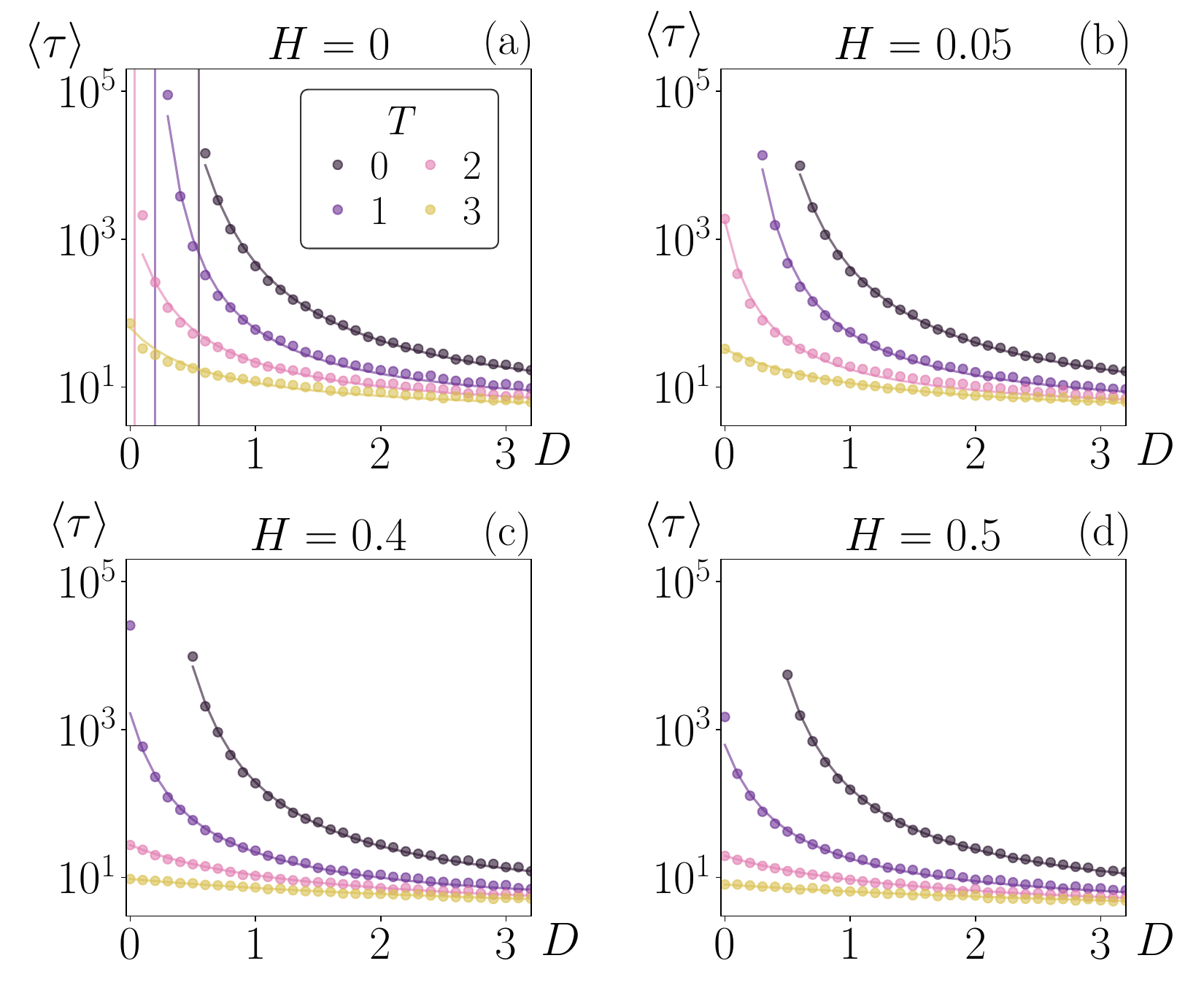}
 \caption{Average first-passage time $\langle\tau\rangle$ from the ordered state $m=-1$ to the barrier region $m=0$, for a system of size $L = 100$, as a function of the random field intensity $D$, for different values of the bias field $H$ and several temperatures. The averages $\langle \tau \rangle$ are computed from a sample of $10^3$ first-passage times. Solid lines correspond to fits of Eq.~\eqref{eq:fit_tau} to the data. Note that all panels share the same vertical scale. In panel (a), vertical lines denote the critical noise intensity $D_0(T)$ separating the broad-ferromagnetic and ferromagnetic phases for $T = 0, 1, 2$, with $D_0(T=0)\approx0.6$, $D_0(T=1)\approx0.2$ and $D_0(T=2)\approx0.04$ (see Fig.~\ref{fig:PhaseDiagramH0}).}
 \label{fig:tauvsDfit_comparison}
\end{figure}

We first discuss the dependence of the average first-passage time $\langle \tau \rangle$ on the noise intensity~$D$ for different values of the bias field $H$. The results for $L=100$ and several temperatures, below and above $\Tc$, are shown in Fig.~\ref{fig:tauvsDfit_comparison}.

In the absence of a bias field ($H=0$) [see Fig.~\ref{fig:tauvsDfit_comparison}(a)], the average first-passage time $\langle \tau \rangle$ decreases by several orders of magnitude as the stochastic field intensity~$D$ increases. For temperatures below the critical value ($T<\Tc$, namely $T = 0, 1, 2$), $\langle \tau \rangle$ grows rapidly upon decreasing~$D$, becoming effectively divergent in the low-noise regime. This behavior reflects the increasing stability of the two symmetric magnetized states and the growth of the barrier separating them. If the low-temperature regime corresponds to a genuine thermodynamic phase, the escape time is expected to diverge at the critical value $D_0(T)$ separating the broad-ferromagnetic and ferromagnetic phases. These thresholds are indicated by vertical lines for $T = 0, 1, 2$. In the thermodynamic limit, this scenario would imply an infinite escape time for $D < D_0(T)$. However, a direct verification of this prediction is beyond the accessible simulation times and system sizes.

This qualitative behavior persists in the presence of a bias field. For $H=0.05$, $H=0.4$, and $H=0.5$ [see panels (a), (b) and (c) of Fig.~\ref{fig:tauvsDfit_comparison}, respectively], the average first-passage time $\langle \tau \rangle$ again decreases monotonically with increasing noise intensity $D$ at all temperatures. The reduction spans several orders of magnitude, indicating that stochastic driving continues to promote barrier crossing even when the symmetry between the two magnetized states is explicitly broken by the bias field. However, the bias qualitatively modifies the low-noise regime. By tilting the effective free-energy landscape toward the preferred state, the bias lowers the escape barrier and can make transitions possible even in the absence of stochastic driving. As a consequence, for sufficiently large fields ($H\gtrsim 0.4$), the escape time remains finite already at $D=0$, although it can still become very large at low temperatures. Except for the lowest temperature $T=0$, this behavior is clearly observed for $H=0.4$ and 0.5. In contrast, for the weaker bias ($H=0.05$), the bias field appears insufficient to fully suppress the metastable barrier at the lowest temperatures, leading to much longer escape times.

Another systematic feature visible in all panels in Fig.~\ref{fig:tauvsDfit_comparison} is the strong temperature dependence of~$\langle \tau \rangle$. At fixed $H$ and $D$, increasing $T$ leads to a reduction of the escape time, consistent with thermally activated dynamics. Nevertheless, the effect of $D$ remains substantial at finite temperature, indicating that stochastic driving acts as an additional activation mechanism rather than simply increasing the thermodynamic effective temperature.

As shown in Fig.~\ref{fig:tauvsDfit_comparison}, numerical results for $\langle \tau \rangle$ as a function of noise intensity~$D$, for all combinations of bias strength~$H$ and temperature~$T$ considered, are well described by an Arrhenius-type expression
\begin{equation}
\langle \tau \rangle = \tau_0 \exp\!\left(\frac{\Delta V}{D + D_1}\right),
\label{eq:fit_tau}
\end{equation}
which provides a useful heuristic description of the data over the accessible range of $D$ and for all values of $H$ considered. Within the barrier-crossing picture, $\Delta V$ may be interpreted as an effective activation barrier for the collective magnetization, while $D+D_1$ acts as an effective noise strength and $\tau_0$ sets the corresponding time scale. However, note that we cannot test the validity of the fit close to the boundary $D_0(T)$ separating the broad-ferromagnetic and ferromagnetic regimes, where the average escape times become extremely large and lie beyond the range accessible to our simulations. The fitted values of $\tau_0$, $\Delta V$, and $D_1$ should therefore be understood as effective parameters characterizing the observed range of escape times.

\begin{figure}[h!]
 \centering \includegraphics[width=1.0\linewidth]{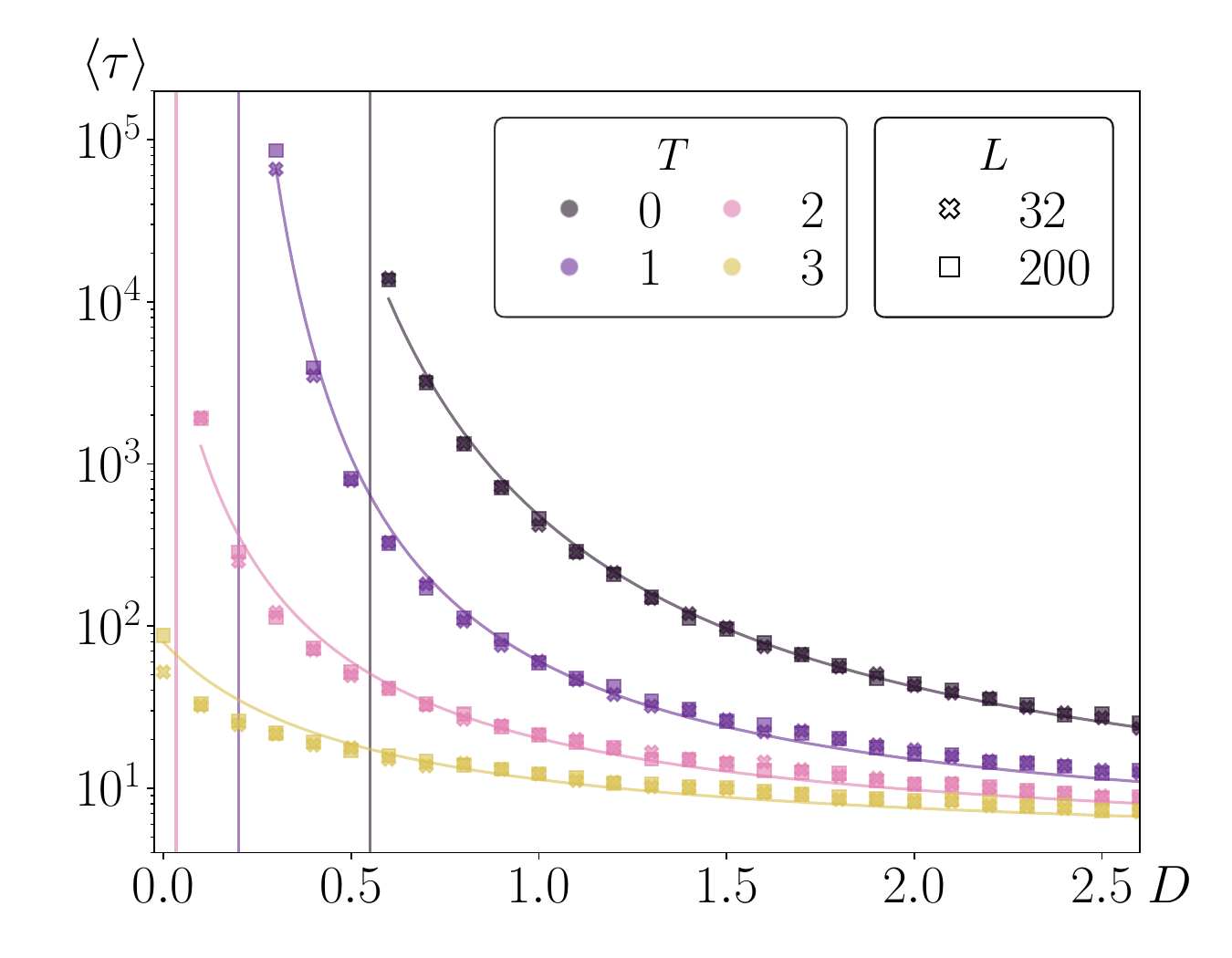}
 \caption {Average first-passage time $\langle\tau\rangle$ as a function of stochastic field strength $D$ for several temperatures and system sizes, in the absence of a bias field ($H = 0$). Note the absence of finite-size effects in the barrier-crossing times. Solid lines show fits of Eq.~\eqref{eq:fit_tau} to the data for $L = 200$. Vertical dashed lines indicate the approximate stochastic intensity separating the broad-ferromagnetic and ferromagnetic phases for \makebox{$T =0,1,2$} (from right to left), with the corresponding values reported in the caption of Fig.~\ref{fig:tauvsDfit_comparison}. All average times $\langle \tau \rangle$ have been calculated from a sample of $10^3$ first-passage times.} 
 \label{fig:tauvsDfit_comparison2}
\end{figure}

A particularly unexpected consequence of stochastic driving is the strong suppression of finite-size effects in the barrier-crossing dynamics. This can be clearly seen in Fig.~\ref{fig:tauvsDfit_comparison2}, where we show the average escape time $\langle \tau \rangle$ as a function of noise intensity $D$ for $H = 0$, considering the same temperatures as in Fig.~\ref{fig:tauvsDfit_comparison} and two system sizes ($L = 32, 200$). In the absence of a bias field, the barrier-crossing process is totally noise-induced, effectively suppressing the finite-size effects that characterize the pure Ising model. The solid lines correspond to fits using the Arrhenius-type expression in Eq.~\eqref{eq:fit_tau}. 

\begin{figure}[h!]
 \centering \includegraphics[width=1.0\linewidth]{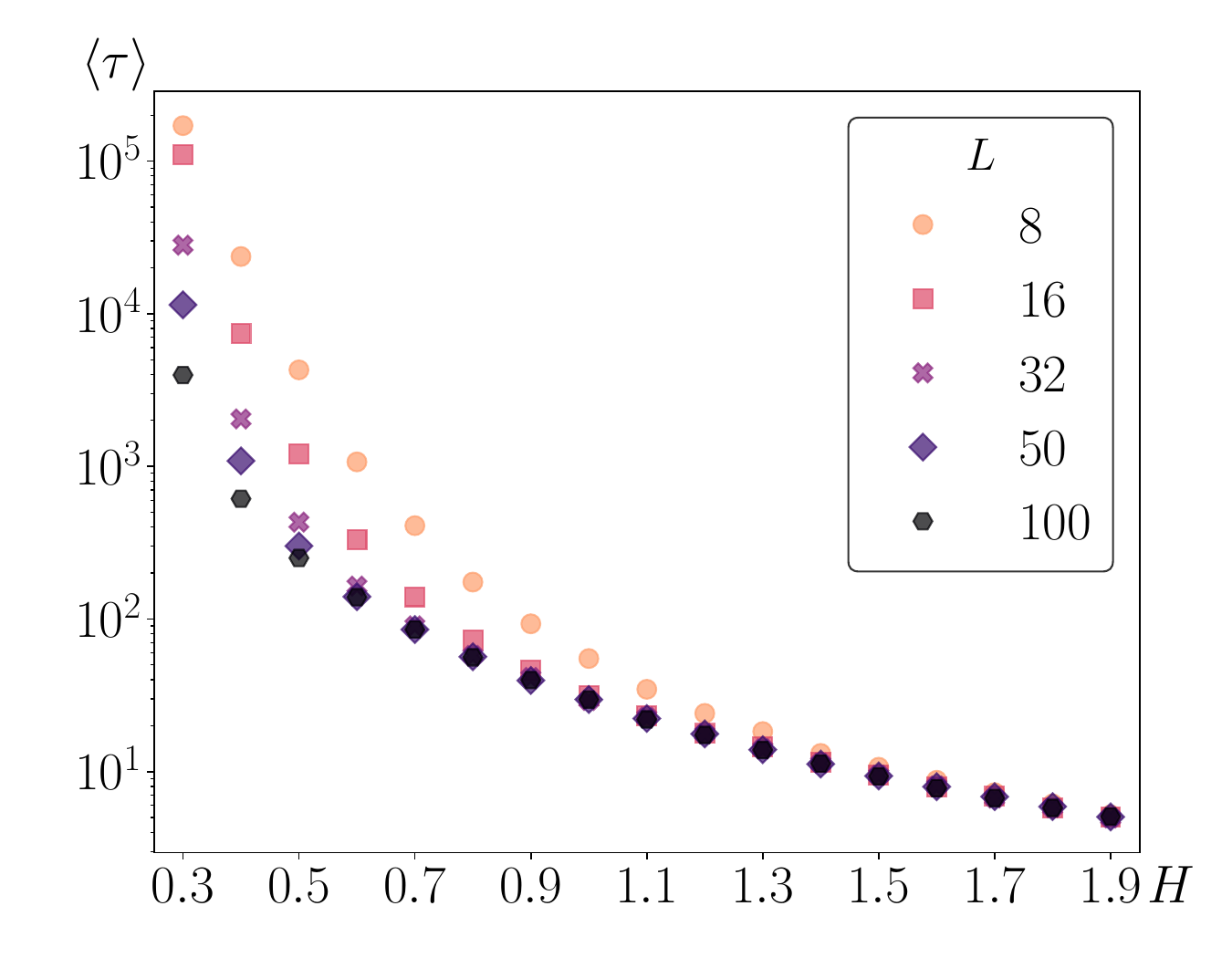}
 \caption{Average first-passage time $\langle\tau\rangle$ as a function of the bias field $H$ for a fixed field intensity $D=0.1$ and several values of the system size $L$. The average times $\langle \tau \rangle$ were computed from $10^3$ independent first-passage events.}
 \label{fig:tauvsH_D01}
\end{figure}

We now examine the dependence of the average escape time $\langle \tau \rangle$ on the bias field $H$ for fixed stochastic intensity $D=0.1$. Fig.~\ref{fig:tauvsH_D01} shows that $\langle \tau \rangle$ decreases steadily with increasing $H$ for all system sizes, spanning several orders of magnitude. This behavior reflects the progressive lowering of the effective barrier as the free-energy landscape is tilted toward the favorable state. 

We therefore conclude that both the bias field and the stochastic field facilitate the barrier-crossing process. Note, however, that excessively large bias fields may substantially perturb the intrinsic properties of the system. Conversely, very strong stochastic driving can induce significant backflow. These results therefore indicate that optimal barrier-crossing performance requires a balanced combination of moderate stochastic driving and controlled bias. 

\section{Conclusions}\label{sec:conclusions}

In this work, we have investigated the combined effects of homogeneous stochastic fields and a weak bias on the barrier-crossing dynamics of a two-state system, represented by the kinetic Ising model. We have found that the bias field primarily determines the location of the dominant maxima of the magnetization distribution $P(m)$, whereas the noise intensity controls their sharpness and overall stability. Additionally, stochastic driving strongly suppresses the system-size dependence of the dynamics.

We have analyzed the switching dynamics from the unfavorable magnetized basin to the favorable one as a function of temperature, noise intensity, system size and bias strength. In the absence of a bias field, the escape process is entirely driven by thermal and stochastic fluctuations. In this regime, the stochastic driving acts as an additional activation mechanism that significantly enhances barrier crossing, causing the escape time $\langle \tau \rangle$ to decrease dramatically with increasing stochastic intensity $D$. The addition of a finite bias field $H$ further reduces the escape time $\langle \tau \rangle$ by tilting the effective free-energy landscape, thereby breaking the underlying symmetry of the bistable potential. The behavior of the escape time is well captured by a generalized Arrhenius form incorporating both thermal and stochastic contributions. 

However, our results indicate that maximizing transition efficiency requires a balanced combination of stochastic driving and bias. For sufficiently large noise intensities $D$, significant backflow into the unfavorable state emerges, which severely reduces directional selectivity. Conversely, an excessively strong bias field $H$ over-perturbs the intrinsic dynamics rather than guiding them efficiently. Consequently, optimal activation is achieved in an intermediate regime, where moderate stochastic driving and a controlled bias act cooperatively to facilitate efficient transitions in bistable systems.

Although the stochastic driving considered here differs from the periodic forcing typically employed in studies of dynamic phase transitions, recent experimental advances in driven magnetic systems suggest possible routes for testing related nonequilibrium barrier-crossing phenomena. In particular, experimental observations of dynamic ordering and anomalous fluctuation regimes in ferromagnetic thin films under external driving~\cite{MarinRamirez2020,Quintana2023} indicate that controlled external fields can produce collective dynamical effects beyond equilibrium behavior. Exploring whether stochastic-field-induced acceleration and symmetry-restoring mechanisms analogous to those described in this paper can be realized in other experimental systems~\cite{wang2020boosted,panta2023quantitatively,martinez2013effective, paneru2023bona,Flanders} remains an interesting direction for future work.\\

\acknowledgments{
A.C. thanks Bret Flanders for many useful discussions. Partial financial support has been received from Grant PID2024-157493NB-C21 funded by MICIU/AEI/10.13039/501100011033 and by ERDF, EU, and the Mar{\'\i}a de Maeztu Program for units of Excellence in R\&D, grant CEX2021-001164-M. S.O-B. acknowledges support from the Spanish Ministry of Education and Professional Training under the grant No. FPU21/04997.}

\bibliography{random_ising}

@article{hromadova2014stochastic,
  title={Stochastic resonance in electron transfer oscillations of extended viologen},
  author={Hromadov{\'a}, Magdal{\'e}na and Val{\'a}\v{s}ek, Michal and Fanelli, Nicolangelo and Randriamahazaka, Hyacinthe N and Pospisil, Lubomir},
  journal={The Journal of Physical Chemistry C},
  volume={118},
  number={17},
  pages={9066--9072},
  year={2014},
   doi = {10.1021/jp501608b},
  publisher={ACS Publications}
}

@article{wang2020boosted,
  title={Boosted molecular mobility during common chemical reactions},
  author={Wang, Huan and Park, Myeonggon and Dong, Ruoyu and Kim, Junyoung and Cho, Yoon-Kyoung and Tlusty, Tsvi and Granick, Steve},
  journal={Science},
  volume={369},
  number={6503},
  pages={537--541},
  year={2020},
  doi = {10.1126/science.aba8425},
  publisher={American Association for the Advancement of Science}
}

@article{panta2023quantitatively,
  title={Quantitatively controlled electrophoretic deposition of nanocrystal films from non-aqueous suspensions},
  author={Panta, Krishna R and Orme, Christine A and Flanders, Bret N},
  journal={Journal of Colloid and Interface Science},
  volume={636},
  pages={363--377},
  year={2023},
  doi = {10.1016/j.jcis.2023.01.004},
  publisher={Elsevier}
}

@article{martinez2013effective,
  title={Effective heating to several thousand kelvins of an optically trapped sphere in a liquid},
  author={Mart{\'\i}nez, Ignacio A and Rold{\'a}n, Edgar and Parrondo, J. M. R.  and Petrov, Dmitri},
  journal={Physical Review E},
  volume={87},
  number={3},
  pages={032159},
  year={2013},
  doi = {10.1103/PhysRevE.87.032159},
  publisher={APS}
}

@article{paneru2023bona,
  title={{Bona fide stochastic resonance under nonGaussian active fluctuations}},
  author={Paneru, Govind and Tlusty, Tsvi and Pak, Hyuk Kyu},
  journal={Soft Matter},
  volume={19},
  number={7},
  pages={1356--1362},
  year={2023},
  doi = {10.1039/D2SM01449A},
  publisher={Royal Society of Chemistry}
}

@article{gammaitoni1998stochastic,
  title={Stochastic resonance},
  author={Gammaitoni, Luca and H{\"a}nggi, Peter and Jung, Peter and Marchesoni, Fabio},
  journal={Reviews of Modern Physics},
  volume={70},
  number={1},
  pages={223},
  year={1998},
  doi = {10.1103/RevModPhys.70.223},
  publisher={APS}
}

@article{acharyya1999nonequilibrium,
  title = {Nonequilibrium phase transition in the kinetic Ising model: Existence of a tricritical point and stochastic resonance},
  author = {Acharyya, Muktish},
  journal = {Phys. Rev. E},
  volume = {59},
  issue = {1},
  pages = {218--221},
  numpages = {0},
  year = {1999},
  month = {Jan},
  publisher = {American Physical Society},
  doi = {10.1103/PhysRevE.59.218},
  url = {https://link.aps.org/doi/10.1103/PhysRevE.59.218}
}

@article{vollmayr1993finite,
  title={Finite size effects at thermally-driven first order phase transitions: A phenomenological theory of the order parameter distribution},
  author={Vollmayr, Katharina and Reger, Joseph D and Scheucher, Manfred and Binder, Kurt},
  journal={Zeitschrift f{\"u}r Physik B Condensed Matter},
  volume={91},
  pages={113--125},
  year={1993},
  doi = {10.1007/BF01316713},
  publisher={Springer}
}

@misc{Flanders,
	author = {B.F. Flanders},
	title = {Private communication}}

@inproceedings{Toral:2011,
   author = {R. Toral},
   doi = {10.1063/1.3577618},
   isbn = {9780735408876},
   issn = {0094243X},
   booktitle = {AIP Conference Proceedings},
   pages = {145-154},
   title = {Noise-induced transitions vs. noise-induced phase transitions},
   volume = {1332},
   year = {2011},
}

@article{BPTK:1997,
   author = {C. {Van den Broeck} and J. M. R. Parrondo and R. Toral and R. Kawai},
   doi = {10.1103/PhysRevE.55.4084},
   issn = {1063-651X},
   issue = {4},
   journal = {Phys. Rev. E},
   month = {4},
   pages = {4084},
   publisher = {American Physical Society},
   title = {Nonequilibrium phase transitions induced by multiplicative noise},
   volume = {55},
   url = {http://pre.aps.org/abstract/PRE/v55/i4/p4084_1 https://link.aps.org/doi/10.1103/PhysRevE.55.4084},
   year = {1997},
}

@article{BPT:1994,
   author = {C. {Van den Broeck} and J.M.R. Parrondo and R. Toral },
   doi = {10.1103/PhysRevLett.73.3395},
   issn = {0031-9007},
   issue = {25},
   journal = {Physical Review Letters},
   month = {12},
   pages = {3395},
   publisher = {APS},
   title = {Noise-Induced Nonequilibrium Phase Transition},
   volume = {73},
   url = {http://prl.aps.org/abstract/PRL/v73/i25/p3395_1 https://link.aps.org/doi/10.1103/PhysRevLett.73.3395},
   year = {1994},
}

@article{Acharyya1998,
   author = {Muktish Acharyya},
   doi = {10.1103/PhysRevE.58.174},
   issn = {1063651X},
   issue = {1},
   journal = {Physical Review E},
   month = {7},
   pages = {174-178},
   publisher = {American Physical Society},
   title = {Nonequilibrium phase transition in the kinetic {Ising} model: Dynamical symmetry breaking by randomly varying magnetic field},
   volume = {58},
   url = {https://link.aps.org/doi/10.1103/PhysRevE.58.174},
   year = {1998},
}

@book{1984Horsthemke,
   author = {W Horsthemke and R Lefever},
   city = {Berlin},
   publisher = {Springer},
   title = {Noise-Induced Transitions},
   year = {1984},
}

@article{tome1990dynamic,
  title={{Dynamic phase transition in the kinetic Ising model under a time-dependent oscillating field}},
  author={Tom{\'e}, T. and de Oliveira, M. J.},
  journal={Physical Review A},
  volume={41},
  number={8},
  pages={4251},
  year={1990},
  publisher={APS},
  url={https://journals.aps.org/pra/abstract/10.1103/PhysRevA.41.4251}
}

@article{lo1990ising,
  title={Ising model in a time-dependent magnetic field},
  author={Lo, W. S. and Pelcovits, R. A.},
  journal={Physical Review A},
  volume={42},
  number={12},
  pages={7471},
  year={1990},
  publisher={APS},
  url={https://journals.aps.org/pra/abstract/10.1103/PhysRevA.42.7471}
}

@article{acharyya1995response,
  title={{Response of Ising systems to oscillating and pulsed fields: Hysteresis, ac, and pulse susceptibility}},
  author={Acharyya, M. and Chakrabarti, B. K.},
  journal={Physical Review B},
  volume={52},
  number={9},
  pages={6550},
  year={1995},
  publisher={APS},
  url={https://journals.aps.org/prb/abstract/10.1103/PhysRevB.52.6550}
}

@article{Yuksel:2022,
    title={Dynamic phase transition in classical Ising models},
    author={Yüksel, Yusuf  and Vatansever, Erol},
    journal={Journal of Physics D: Applied Physics},
    year={2022},
    volume={55},
    issue={7},
    pages={073002},
    doi = {10.1088/1361-6463/ac2f6c}
}

@article{Li_2024,
  title = {Dynamic magnetic characteristics of the kinetic Ising model under the influence of randomness},
  author = {Li, {Bo-chen} and Wang, Wei},
  journal = {Phys. Rev. E},
  volume = {110},
  issue = {3},
  pages = {034134},
  numpages = {14},
  year = {2024},
  month = {Sep},
  publisher = {American Physical Society},
  doi = {10.1103/PhysRevE.110.034134},
  url = {https://link.aps.org/doi/10.1103/PhysRevE.110.034134}
}

@article{Ghosh_2013,
doi = {10.1088/1742-5468/2013/11/P11015},
url = {https://doi.org/10.1088/1742-5468/2013/11/P11015},
year = {2013},
month = {nov},
publisher = {IOP Publishing and SISSA},
volume = {2013},
number = {11},
pages = {P11015},
author = {Ghosh, Asim and Chakrabarti, Bikas K},
title = {Response of the two-dimensional kinetic Ising model under a stochastic field},
journal = {Journal of Statistical Mechanics: Theory and Experiment},
}

@article{paper1,
  title = {Stochastic field effects in a two-state system: Symmetry breaking and symmetry restoring},
  author = {Oliver-Bonafoux, Sara and Toral, Ra\'ul and Chakrabarti, Amitabha},
  journal = {Phys. Rev. E},
  volume = {113},
  issue = {4},
  pages = {044110},
  numpages = {10},
  year = {2026},
  month = {Apr},
  publisher = {American Physical Society},
  doi = {10.1103/kyxg-y54l},
  url = {https://link.aps.org/doi/10.1103/kyxg-y54l}
}

@article{Baxter2011,
	author = {Baxter, R.  J.},
	date = {2011/11/01},
	doi = {10.1007/s10955-011-0213-z},
	id = {Baxter2011},
	isbn = {1572-9613},
	journal = {Journal of Statistical Physics},
	number = {3},
	pages = {518--548},
	title = {Onsager and {K}aufman's {C}alculation of the {S}pontaneous {M}agnetization of the {I}sing {M}odel},
	url = {https://doi.org/10.1007/s10955-011-0213-z},
	volume = {145},
	year = {2011}}

@article{onsager1944crystal,
  title={{Crystal Statistics. I. A Two-Dimensional Model with an Order-Disorder Transition}},
  author={Onsager, L.},
  journal={Physical Review},
  volume={65},
  number={3-4},
  pages={117},
  year={1944},
  publisher={APS},
  url={https://journals.aps.org/pr/abstract/10.1103/PhysRev.65.117}
}

@book{Newman2023,
   author = {M. E. J. Newman and G. T. Barkema},
   isbn = {9780198517979},
   pages = {475},
   publisher = {Clarendon Press},
   title = {Monte Carlo Methods in Statistical Physics},
   url = {https://global.oup.com/academic/product/monte-carlo-methods-in-statistical-physics-9780198517979?cc=es&lang=en&},
   year = {2023},
}

@article{Quintana2023,
  title = {Metamagnetic fluctuation characteristics near dynamic phase transitions},
  author = {Quintana, M. and Mart\'{\i}n Valderrama, C. and Berger, A.},
  journal = {Phys. Rev. E},
  volume = {108},
  issue = {6},
  pages = {064121},
  numpages = {9},
  year = {2023},
  month = {Dec},
  publisher = {American Physical Society},
  doi = {10.1103/PhysRevE.108.064121},
  url = {https://link.aps.org/doi/10.1103/PhysRevE.108.064121}
}

@article{MarinRamirez2020,
  title = {Experimental exploration of dynamic phase transitions and associated metamagnetic fluctuations for materials with different Curie temperatures},
  author = {Mar\'{\i}n Ram\'{\i}rez, J. M. and Oblak, E. and Riego, P. and Campillo, G. and Osorio, J. and Arnache, O. and Berger, A.},
  journal = {Phys. Rev. E},
  volume = {102},
  issue = {2},
  pages = {022804},
  numpages = {12},
  year = {2020},
  month = {Aug},
  publisher = {American Physical Society},
  doi = {10.1103/PhysRevE.102.022804},
  url = {https://link.aps.org/doi/10.1103/PhysRevE.102.022804}
}

\end{document}